# A quantification of how much crypto-miners are driving up the wholesale cost of energy in Texas

**Jangho Lee, Lily Wu, and Andrew E. Dessler**

Department of Atmospheric Sciences, Texas A&M University, College Station, TX, USA

Corresponding author: Andrew Dessler (adessler@tamu.edu)

**Abstract**

The use of energy by cryptocurrency mining comes not just with an environmental cost but also an economic one through increases in electricity prices for other consumers. Here we investigate the increase in wholesale price on Texas' ERCOT grid due to energy consumption from cryptocurrency mining. For every GW of cryptocurrency mining load on the grid, we find that the wholesale price of electricity on the ERCOT grid increases by 2%. Given that today's cryptocurrency mining load on the ERCOT grid is around 1 GW, it suggests that wholesale prices have already risen this amount. There are 27 GW of mining load waiting to be hooked up to the ERCOT grid. If cryptocurrency mining increases rapidly, the price of energy in Texas could skyrocket.

# 1. Introduction

One of the most widely discussed aspects of mining of cryptocurrencies (hereafter, crypto mining) is its enormous energy footprint. This energy consumption is fundamental to the security of so-called proof-of-work coins, where decisions are made by consensus of the mining network. As more and more people join the bitcoin mining network and it consumes more and more power, it becomes more secure (1, 2). At this point, crypto mining's energy consumption is astronomical, consuming as much energy as a small country (3, 4, 5).

This enormous energy use comes with an environmental cost. If the electricity being consumed comes from fossil fuels, then this mining drives climate change; this has become one of the biggest criticisms of proof-of-work cryptocurrencies (6, 7, 8, 9). Less well appreciated is its economic cost: consumption of energy by crypto-miners translates into higher prices for everyone else. This was discussed in Benetton, Compiani (10), who estimated that bitcoin raised energy prices for small businesses and households in Upstate NY and, in China, hurt wages and investments. This paper examines the impact of crypto mining on electricity prices within the Electric Reliability Council of Texas (ERCOT) grid. ERCOT is an independent system operator that manages the distribution of electricity throughout the state of Texas. Our analysis aims to determine the extent to which the energy consumption associated with crypto mining has contributed to increases in ERCOT's wholesale electricity prices.

# 2. Demand-Price Relationship

We analyze the period from March 2021 to August 2022. We calculate two different versions of demand: total hourly demand, as provided directly from ERCOT (https://www.ercot.com/gridinfo/generation) and hourly net demand, which we calculate as the total demand minus wind and solar power production, as was done in Rhodes (11). For the price data, we also calculate two versions: hourly hub-averaged real-time wholesale market price (https://www.ercot.com/mp/data-products/markets/real-time-market?id=NP6-785-ER) and hourly hub-averaged day-ahead wholesale market prices (https://www.ercot.com/mp/data-products/markets/day-ahead-market?id=NP4-180-ER).

We then develop a demand-price curve for each year (2021, 2022) for each combination of demand (total demand, net demand) and price (day-ahead price, real-time price). To do this, we bin the data into 2.5-GW demand bins and calculate the median price in each bin. We then take the derivative of this relationship and smooth that with a 3-bin running average.

Fig. 1 shows hourly total demand vs. day-ahead wholesale price for 2022. As total demand increases, so does the price, reflecting the fact that ERCOT uses a single clearing price for the wholesale market, where that clearing price is set by the most expensive energy source that is required to satisfy the demand. With higher demand, ERCOT must turn to progressively more expensive generators, leading to the increase in price (12, 13). The results are similar when we use real-time price (the dashed line in Fig. 1). We have also analyzed net demand, which is total demand minus solar and wind power, and it also shows similar behavior (not shown).

The total demand-price relationship is generally linear until the price abruptly increases above ~70 GW of total demand. This "dogleg" in the demand-price curve is caused by congestion on the electrical grid as well as the grid nearing its capacity, both of which lead to rapid increases in prices.

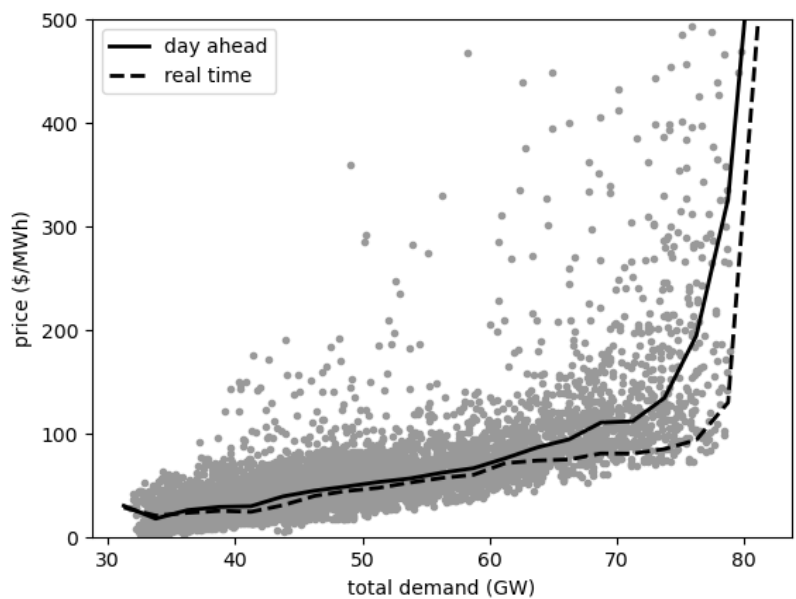

**Figure 1.** Relationship between hourly total demand and wholesale day-ahead price. Dots show the hourly values from 2022. The solid line shows the median of day-ahead price; the dashed line shows the median value using real-time price (hourly real-time prices not shown).

### 3. Increase of Electricity Price due to Crypto Mining

Figure 2 shows the derivative of the demand-price curve in Fig. 1. This is the increase in price per GW increase in demand (see methods for details of how this is calculated). From these derivatives, we estimate the overall average price increase as demand increases, as described in the methods. This yields an increase in the wholesale price due to increasing demand of 1.6-2.4% per GW (Table 1).

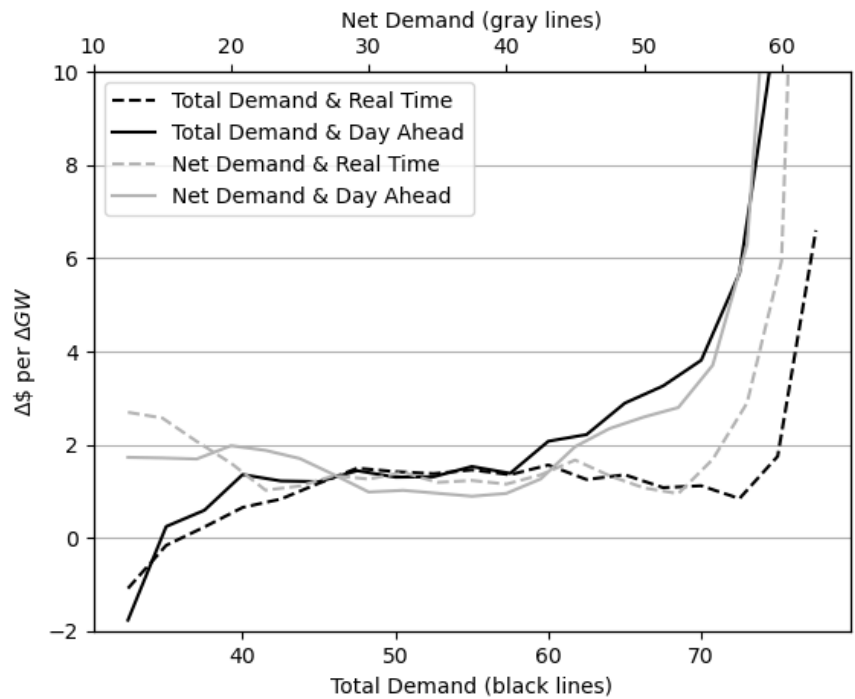

**Figure 2**. Increase in wholesale price per GW increase in electrical demand, as a function of demand. The four curves represent the four combinations of the two price and two demand data sets. The black lines on the plot represent the values for total demand and are plotted along the bottom x-axis. The gray lines on the plot represent the values for net demand and are plotted along the top x-axis. Solid lines use day-ahead prices and dashed lines use real-time prices.

**Table 1**. Increase in wholesale prices in % due to increased demand

|  | 1 GW | 5 GW |
|---|---|---|
| total demand + day-ahead price | 2.2% | 11.2% |
| total demand + real-tme price | 1.6% | 8.2% |
| net demand + day-ahead price | 2.4% | 11.8% |
| net demand + real-tme price | 2.0% | 10.2% |

Presently, crypto mining consumes about 1 GW of power from the ERCOT grid (14). To calculate the price increase due to crypto mining, we take the hourly demand and the derivative of price with respect to demand (Fig. 2) and calculate the hourly increase in price for a 1- and 5-GW increase in hourly demand. We then average this time series, with each hour weighted by demand. We then express this as a fraction of the demand-weighted average hourly wholesale price of power (Table 1).

Cryptocurrency miners may curtail their electricity usage when the electricity demand is high. There is little public information about when they do this, so we assume that it occurs when demand exceeds 70 GW of total demand or 55 GW of net demand, consistent with the "doglegs" in Fig. 1. Crypto miners get paid for this demand curtailment, which can amount to 10% of the income of a miner (15, 16, 17) but we do not include the costs of these grid services in our analysis of the costs of crypto mining.

While crypto mining has increased the wholesale price of electricity by about 2% today (Table 1), a potentially bigger concern is the future: According to ERCOT, there is 27 GW of crypto-mining load waiting to be connected to the Texas grid (14, 18). Even an increase in demand of a few GW would significantly increase the wholesale cost of electricity on the Texas grid — for example, 5 GW would increase the wholesale price by 8-12% (Table 1), imposing a significant economic burden on Texas consumers.

We should note one caveat to this: the feedback between price and demand. An increase in the wholesale price provides the market with a signal to encourage the construction of low-cost power (like wind and solar). Thus, increases in demand from crypto mining, and the accompanying increase in price, should incentivize construction of additional low-cost generation, which in turn should eventually decrease the price.

This feedback, however, has a time scale: building new generation capacity is not instantaneous, but takes several years or more. Because the increase in energy consumption in Texas due to crypto mining has occurred rapidly, in the last year or two, the market has not adjusted to the increased demand. This justifies our assumption that the increase in demand from crypto mining has driven an increase in price that has not been offset by a corresponding change in price-demand curve.

Residential consumers tend to be on fixed price plans whose prices change only occasionally. However, we expect increases in wholesale price to ripple through the economy and raise energy costs for everyone, including residential consumers. Figure 3 shows that residential prices are indeed strongly correlated with wholesale prices (https://www.eia.gov/electricity/sales_revenue_price/), although the magnitude of the increase in residential price is 39% of the increase in wholesale price.

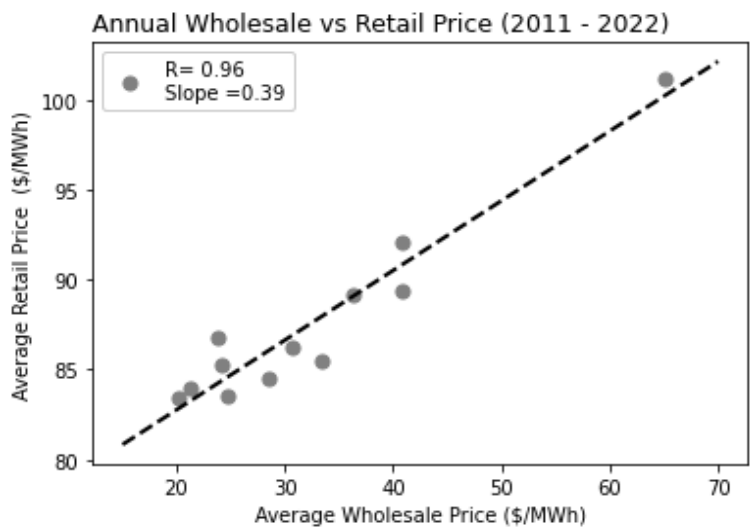

**Figure 3.** Comparison between annual averaged wholesale price (hub price) and retail price for 2011-2022 period. Each dots represents the price for each year, and dashed line shows the linear regression line between wholesale and retail price.

**4. Discussion**

We see strong evidence of a "bitcoin tax" on the Texas economy. We estimate that it has already raised wholesale energy prices by about 2% and the proposed rapid expansion of crypto mining in Texas could drive much larger future increases. If Texans want to expand crypto mining without causing economically painful increases in the price of electricity, there there must be a corresponding build-out of low-cost generation capacity. This can occur with specific government policies, such as requiring crypto miners to build low-cost generation, or by allowing new crypto mining load to connect to the grid slowly enough that the market responds by itself.

*Acknowledgments:* This work was supported by NSF grant AGS-1841308 to Texas A&M University. We thank Dr. Joshua Rhodes for useful discussions.